\documentclass[sn-mathphys,Numbered]{sn-jnl}

\usepackage[title]{appendix}
\usepackage{amsmath}
\usepackage{amssymb}
\usepackage{amsthm}
\usepackage{booktabs}
\usepackage{color}
\usepackage{dsfont}
\usepackage{epstopdf}
\usepackage{float}
\usepackage{graphicx}
\usepackage{mathptmx}
\usepackage{subcaption}
\usepackage{tabularx}
\usepackage{textcomp}
\usepackage{algorithm}
\usepackage{titlesec}
\usepackage{array,multirow}%
\usepackage{mathrsfs}%
\usepackage{adjustbox}
\usepackage{textcomp}%
\usepackage{mathtools}
\usepackage{manyfoot}%
\usepackage{hyperref}
\makeatletter
\renewcommand{\ALG@name}{Process}
\makeatother
\usepackage{algorithmicx}%
\usepackage{algpseudocode}%
\usepackage{listings}%
\usepackage{makecell}
\usepackage[table]{xcolor}
\usepackage{placeins}
\usepackage{subcaption}

\newcommand{\agent}{i}
\newcommand{\nagents}{\textbf{I}}
\newcommand{\tableno}{j}
\newcommand{\ntables}{\textbf{J}}
\newcommand{\roundno}{k}
\newcommand{\nrounds}{K}

\newcommand{\groupselect}{\textsc{Group\-Se\-lect}}

\newcommand{\legacy}{\textsc{Legacy}}
\newcommand{\groupopt}{\textsc{Dream}}
	
\usepackage{lineno}

\begin{document}

\title[A New Heuristic Algorithm for Balanced Deliberation Groups]{A New Heuristic Algorithm for Balanced Deliberation Groups}


\author*[1]{\fnm{Jake} \sur{Barrett}}\email{j.t.l.barrett@sms.ed.ac.uk}

\author[2]{\fnm{Philipp C.} \sur{Verpoort}}\email{pcv@sortitionfoundation.org}

\author[3,1]{\fnm{Kobi} \sur{Gal}}\email{kgal@ed.ac.uk}

 \affil*[1]{\orgdiv{School of Informatics}, \orgname{University of Edinburgh}, \orgaddress{\street{10 Crichton Street}, \city{Edinburgh}, \postcode{EH8 9AB}, \state{Edinburgh}, \country{Scotland}}}

\affil[2]{\orgname{Sortition Foundation, Cambridge, UK}}

\affil[3]{\orgdiv{Department of Software and Information Systems Engineering}, \orgname{Ben-Gurion University}, \orgaddress{\city{Be'er Sheva}, \country{Israel}}}


\abstract{We here present an improved version of the Sortition Foundation's \groupselect{} software package, which aims to repeatedly allocate participants of a deliberative process to discussion groups in a way that balances demographics in each group and maximises distinct meetings over time. Our result, \groupopt{}, significantly outperforms the prior algorithmic approach \legacy{}. We also add functionalities to the \groupselect{} software to help the end user. The \groupopt{} algorithm utilises random shuffles and Pareto swaps to find a locally optimal solution that maximises demographic balance and minimises the number of pairwise previous meetings, with the relative importance of these two metrics defined by the user.
}




\maketitle

\section{Introduction}
Deliberative processes, in their narrow sense as a practice in the context of deliberative techniques, have gained substantial interest from academics, policymakers, practitioners, and advocates in recent years \cite{Flanigan2021FairAssemblies,Flanigan2020NeutralizingSortition,Curato2023GlobalReport,Boswell2023IntegratingAssemblies,Andrews2022ScotlandsExperience}. They involve anything from less than ten to several hundreds of participants who are selected randomly or in a targeted fashion. Deliberative processes may last anywhere between a few hours and several days, and may be conducted in-person, online, or in a hybrid combination. When participants are selected randomly from the wider population (a practice also known as `sortition'), these processes are also referred to as deliberative mini-publics and are commonly known as `Citizens' Assemblies', `Citizens' Juries', or `Citizens' Panels' \cite{Escobar2017FormsPractice}. Deliberative processes are often focused on specific topics, but may also be completely open to topics brought up by participants. Moreover, deliberative processes are usually facilitated by trained independent individuals and follow specific conversation rules that guard the deliberative nature. For a more detailed introduction to deliberative processes and techniques, we refer readers to further material from practitioners and institutional bodies \cite{OECD2020InnovativeInstitutions,Escobar2017FormsPractice}.

A crucial component of deliberative processes are open verbal facilitated exchanges in small groups of typically 5--8 (but not more than 10) individuals. With many deliberative processes involving a large multiple of those numbers, there is a need to subdivide the group of all participants into smaller groups at least once in a process. This subdivision may also be repeated with different group allocations, depending on the time frame of the process. Moreover, some past deliberative processes split participants into themed groups to focus on subtopics (e.g. splitting climate action into energy supply, transport, and housing) in order to be able to dive more deeply into those subtopics within a limited time frame \cite{Andrews2022ScotlandsExperience}.

When subdividing participants of deliberative processes into smaller groups, practitioners have a range of options: (a) they could allocate participants to groups randomly, (b) they could allow participants to decide those allocations themselves, or (c) they could optimise those allocations for desired properties. Throughout the rest of this article, we focus on the last option.

In that case, practitioners would typically optimise their group allocations regarding one or more of these goals:
\begin{enumerate}
    \item Maximise diversity of specific features of participants in each group (e.g. mix of genders, ages, regional backgrounds, etc).
    \item Cluster specific features between groups (often due to technical requirements such as there only being one room, one table, or one facilitator being able to cater to special needs of participants such as disabilities, language translation, or media consent).
    \item Maximise the number of distinct meetings between individuals across different allocations.
\end{enumerate}
Target (1) may be important to ensure that the diversity of the full group of participants is also present in each smaller group. This is designed to maximise epistemic diversity, ensuring that ``different social styles and demographics [are distributed] evenly'' \cite{newDemocracyFoundation2018EnablingElections}. As an example of such demographic mixing, the second weekend of the Citizens' Assembly of Scotland diversified across three demographic variables: ``tables were sorted Saturday by political view, age and gender. Tables on Sunday sorted by political view, gender then age'' \cite{Elstub2022ResearchScotland}. Through conversations with facilitators, the motivation of target (2)  is mainly to ensure that individuals who do not consent to media recording are grouped together; for finer-grain clustering, manual allocations may be required (e.g. ``artificially place the hearing-impaired participants on table 4, with an interpreter, and optimise everything else around this''). Target (3) may be important to ensure that participants get exposed to the views and perspectives of as many other participants as possible and to avoid that link-minded individuals forming cliques with like-minded individuals and creating identities that hinder critical thinking; mixing helps with the goal of ``find[ing] common ground across the whole diverse group'' \cite{newDemocracyFoundation2018EnablingElections}. 

One could however argue against the optimisation of allocations w.r.t. to goals (1), (2), and (3) based on the fact that the features that would be important to diversify may be hard to determine (known unknowns) or may have not come to the understanding of those facilitating the process (yet). Moreover, goals (1), (2), and (3) may oppose each other. For instance, trying to optimise the diversity in group allocations across too many dimensions could constrain the set of possible allocations very strongly, resulting in few distinct meetings between participants. Here, we do not argue for one or another way of optimising allocations, but leave it to practitioners themselves to decide which goals they would like to optimise.

In order to facilitate such an optimisation at low effort and with high performance according to the set targets, the software \groupselect{} was created in 2020 by the Sortition Foundation (an international not-for-profit organisation headquartered in the UK that promotes the use of sortition and deliberation and delivers deliberative processes convened by public authorities on a regular basis). This software packages a heuristic allocation algorithm, here referred to as \legacy{}. Using the \legacy{} algorithm, the software gives users the option to allocate all participants of a given deliberative process into any (feasible) number of groups and corresponding size and do so repeatedly for several different setups.

Theoretical attempts to improve upon \legacy{} introduced in Barrett et al \cite{Barrett2023NowOptimization} rely on the cutting-edge Gurobi optimisation solver, which is free for academic licenses but prohibitively expensive for industry applications.\footnote{\url{https://www.gurobi.com/request-a-price-quote/} (accessed 22/07/2024). We are aware of a non-profit Gurobi track for \href{https://www.gurobi.com/company/gurobi-gives-back/}{non-profit use}, which is for ``qualifying non-profit organisations working to support at least one of the United Nations' Sustainable Development Goals''. However, we see this as insufficient for our purposes as we cannot guarantee that all end-users would satisfy this criteria, particularly in light of their clarification that a ``government agency requiring Gurobi for operations'' does not count as non-profit, and government agencies are significant drivers of large-scale CAs} While open-source solvers have improved in recent years, they still fail on realistic table allocation use cases due to the exponential parameter growth \cite{Huangfu2018ParallelizingMethod}. Therefore, we require a heuristic approach that leverages the insights from an optimised approach and outperforms the current art.

In this work we introduce a way to achieve the above goals of a good allocation algorithm: (1) balanced demographics, (2) a clustering option, and (3) maximal unique pairs. In previous investigations of the \legacy{} algorithm \cite{Barrett2023NowOptimization}, it was found that the approach struggled to satisfy objective (3), particularly over large numbers of rounds. This was because the heuristic approach is effectively deterministic up to swaps within sets of demographically identical individuals. Furthermore, the \legacy{} algorithm only considers first meetings between pairs, with no preference over subsequent meetings: so a pair meeting 10 times is considered equally as valuable to a process as a pair meeting twice. 

Our new approach utilises random shuffles and local Pareto optimisation to attain a stronger meeting profile than that achieved by \legacy \   \cite{SortitionFoundation2021GroupAlgorithm}.  The updated version of the \groupselect{} app, containing the improved heuristic algorithm \groupopt{} (Diverse Representation and Engagement Allocation Method), is available \href{https://github.com/jake126/groupbalance-app/tree/jb-edit}{here}.

\section{Notation}
\label{sec:setting}

We define a setting of $\nagents$ agents in the deliberative panel, $\ntables$ discussion groups (henceforth referred to as `tables', even if they represent an online space), and $\nrounds$ rounds of discussion. In each round $\roundno \in \nrounds$, each agent $\agent \in \nagents$ is allocated to  table $\tableno \in \ntables$, using an allocation function $A_\roundno: \nagents \rightarrow \ntables$. Further notation is specific to the \groupopt{} approach:

\begin{itemize}
    \item Agents $\agent_m \in \nagents_m$ are manually allocated to tables
    \item A subset of tables $\ntables^c$ are reserved for individuals with clustering requirements 
    \item Each demographic $d \in D$ has a set of possible values $\{V_d\}$, and for each value $v_d \in \{V_d\}$, we define the proportion of the panel with demographic $d=v_d$ as $p_{v_d}^P:=\frac{\sum_\agent d(\agent)=v_d}{ \mid\nagents\mid}$, where $d(\agent)$ is the value of demographic $d$ for participant $\agent$. Similarly, we define $p_{v_d,\roundno}^j:= \frac{\sum_{\agent:A_\roundno(\agent)=j} d(\agent)=v_d}{\sum_\agent 1_{A_\roundno(\agent)=j}}$ as the proportion of agents on table $\tableno$ with demographic $d=v_d$ in round $\roundno$
    \item For a specific allocation profile $A_\roundno$, if agent $\agent_1$ on table $\tableno_1$ were swapped with agent $\agent_2$ on table $\tableno_2$, we can define the resulting proportion on table $\tableno_1=A_\roundno(\agent_1)$ as $$p_{v_d,\roundno}^{\roundno(\agent_1 \rightarrow i_2)}:= \frac{\sum_{\agent \neq i_1:A_\roundno(\agent)=j_1,\agent_2} d(\agent)=v_d}{\sum_\agent 1_{A_\roundno(\agent)=j}}$$
    \item The distance of each table $\tableno$ from the overall panel proportion for demographic $d$ in round $\roundno$ is defined as $$\Delta_{\tableno,d,\roundno}:=\sum_{v_d \in V_d}||p_{v_d,\roundno}^j-p_{v_d}^P||$$
\end{itemize}

In an ideal allocation we aim for $p_{v_d,\roundno}^j=p_{v_d}^P$ $\forall j$, $\forall v_d \in V_d$, $\forall d \in D$; therefore, the algorithmic approach aims to minimize $\Delta_{\tableno,d,\roundno}$ for all tables $\tableno$ and all demographics $d$. Note that distances following a swapping of agents $\agent_1,\agent_2$ can be analogously defined using the notation $\tableno(\agent_1 \rightarrow i_2).$ We can also aggregate distances across tables and rounds to gain a measure of the demographic balance of the whole allocation $A$: $\bar\Delta^A:=mean_{\tableno,d,\roundno}(\Delta^A_{\tableno,d,\roundno})$.

\section{Methodology and Heuristic Algorithm}
\label{sec:app_methods}

The \groupopt{} algorithm is described in Algorithm \ref{alg:groupbalance}:

\begin{algorithm}
\caption{\groupopt{} Algorithm}\label{alg:groupbalance}
\begin{algorithmic}[1]
\Require Number of swap rounds $n_S$
\For{$\roundno \in \{1...\nrounds\}$}:
\State Allocate manually allocated agents in $I_m$
\State Randomly allocate clustered agents on a subset of cluster tables $\ntables^c$
\State Randomly allocate non-clustered agents among remaining spaces 
\State Perform $n_S$ rounds of Pareto swaps
\EndFor
\end{algorithmic}
\end{algorithm}

Note that this approach ensures that clustering is automatically satisfied. The user can increase the number of clustering tables to give looser constraints and better performance depending on their priorities. We define the Pareto change for demographic $d$ of swapping individual $\agent$ with individual $\agent'$ for table $\tableno$ as: 

$$P(\agent,\agent',\tableno,d,\roundno) = sign(\Delta_{\tableno,d,\roundno}-\Delta_{\tableno(\agent\rightarrow i'),d,\roundno})$$

If the swap improves/decreases the demographic distance for table $\tableno$, it gets score 1: if there is no change, score 0, and if it worsens/increases the demographic distance, it gets score -1. A Pareto improvement means that every diversifying demographic in the swap is either the same as original or represents an improvement for the table, so $P(\agent,\agent',\tableno^*,d) \geq 0$ $\forall d$, $\forall j^*\in\{A_\roundno(\agent),A_\roundno(\agent')\}$. 

For example, for binary demographic $d=$age and value $v_d=$young, if the age split is 50/50 in the panel and we are on a table with 2 young and 8 old, then $p_{v_d,\roundno}^j=0.2$ and $\Delta_{\tableno,d,\roundno}=0.6$ (the distance from optimal for old + young). We can swap one of the 8 old people for anyone, as both options do not Pareto harm the table for age, or a young person for a young person, but swapping a young person for an old person would push us further from 50/50 so would not be allowed. Therefore, $P(\agent,\agent',\tableno,d,\roundno)=1$ if $d(\agent)=$ old and $d(\agent')=$ is young, $P(\agent,\agent',\tableno,d,\roundno)=0$ if $d(\agent)=d(\agent')$, and $P(\agent,\agent',\tableno,d,\roundno)=-1$ if $d(\agent)=$young, $(\agent')=$old.

For each swap $S$, we can aggregate Pareto changes into a Pareto score $P_S$, which is defined as:

\begin{equation*}
P_S(\agent,\agent',\tableno,\roundno) = \begin{cases}
-1 & any(P(\agent,\agent',\tableno,d,\roundno)=-1)\\
\sum_d P(\agent,\agent',\tableno,d,\roundno) &\text{otherwise}
\end{cases}
\end{equation*}

Pareto swaps are consequently defined in Algorithm \ref{alg:paretoswap}. The process can thus be described as follows: collect the set of swaps that Pareto improve the situation, then with probability $p_S$ choose a swap proportionate to the Pareto score, or with probability $1-p_S$ choose a swap proportionate to the difference in previous meetings across both tables pre- and post-swap. Note that none of the candidate swaps reduce Pareto scores, while some may reduce meeting scores, so $p_S$=0.5 still prioritises demographics over meetings.

\begin{algorithm}
\caption{Pareto Swaps}\label{alg:paretoswap}
\begin{algorithmic}[1]
\Require Pareto swap parameter $p_S$
\For{$\agent \in \{1... \mid\nagents\mid\}$}
\If{$\agent \notin \nagents_m$}
    \State Identify set of potential swaps $\nagents' = \{\agent' \neq i: (P_S(\agent,\agent',\tableno,\roundno) \geq 0$) \& (clustering constraints not violated by the swap)$\}$
    \State Evaluate each swap by its effect on the meeting score: $\{(m_{\tableno,\roundno}-m_{\tableno(\agent\rightarrow i'),\roundno})+m_{\tableno',\roundno}-m_{\tableno'(\agent'\rightarrow i),\roundno})\}$
    \State Eliminate swaps that are strictly dominated by other swaps (i.e. have a worse or equal Pareto score AND a worse or equal meeting score)
    \State Using global (user input) parameter $p_S$ (initialised at 0.5), choose a swap that is proportional to the Pareto score with probability $p_S$ and a probability that is proportional to (non-negative) meeting score improvements with probability (1-$p_S$)
\Else{}
    \State Perform no swaps
\EndIf
\EndFor
\end{algorithmic}
\end{algorithm}

\subsection{Algorithmic Implementation}
The code implementation of \groupopt{} is packaged with an updated version of the \groupselect{} interface with some additional error messages and user guidance around input variables. In particular, interface improvements are:

\begin{itemize}
    \item An improved array of error messages, with suggested parameter inputs if constraints on clustering are tight.
    \item A more accurate calculation of the maximum possible number of links based on the size of the panel and tables, as shown in Barrett et al \cite{Barrett2023NowOptimization}. The previous interface gave a na\"ive maximum calculation based on the combinatoric number of pairs in the panel.
    \item Clustering input updates: as clustering is mainly used to cluster correlated variables (namely consenting to media and conversational tracking), the option to cluster these variables separately is inefficient and limits \legacy{}'s performance in other areas. It makes more sense to collect the clustering constraints into a single binary variable, where a single value in the variable denotes that individuals with this value should be clustered together. The algorithm ensures that all clustering constraints are satisfied before moving on to diversification considerations. 
    \item User chooses number of tables and number of clustering tables, rather than number of tables and number of seats per table. This reduces the cognitive burden on the user, and gives more flexibility in clustering (the previous implementation filled tables until the clustering constraint was filled, leading to artificially tight constraints).
\end{itemize}

\subsection{Experiment Set-up}
Each experiment uses a unique combination of the following inputs:
\begin{itemize}
   \item Data sets (9 permutations):\footnote{The data sets here correspond exactly to \texttt{hd} (size 30), \texttt{sf\_f} (size 40), and extrapolated versions of \texttt{hd} with proportionately representative features (size 100 or 120), taken from experiments from Barrett et al \cite{Barrett2023NowOptimization}}
   \begin{itemize}
     \item $ \mid\nagents\mid=30$, $n_d=3$ demographics, of which 1 can be used for clustering. Each variable is binary in this data set
     \item $ \mid\nagents\mid=40$, $n_d=7$ demographics, of which 1 can be used for clustering (note that an extra experiment was run using only 4 of the demographics for diversification). Four variables are binary (of which one is used for clustering), one has three levels, one has four levels, and the last variable has five possible values
     \item $ \mid\nagents\mid=100$, $n_d=5$ demographics, of which 1 can be used for clustering. Four of the variables are binary (of which one is used for clustering), while the last has four possible values 
     \item $\mid\nagents\mid=120$, $n_d=4$ demographics. Variable options are the same as in the $\mid\nagents\mid=100$ case, excluding the clustering variable
   \end{itemize}
   \item Number of tables (3 permutations): $\mid\ntables\mid$,$\mid\ntables\mid-1$,$\mid\ntables\mid+1$ for whichever $\mid\ntables\mid$ gives closest to 10 members on a table. Note that for the $\mid\nagents\mid=120$ case, to reflect cases where large numbers of participants are divided into thematic subgroups that focus on different topics \cite{Andrews2022ScotlandsExperience}, this experiment is run with $\mid\ntables\mid$=3
   \item Number of rounds (3 permutations): $\nrounds \in \{3,5,10\}$
\end{itemize}

For each experiment with a clustering variable, in \groupopt{}, we used a binding number of cluster tables as this is most closely comparable to \legacy{}. For example, if tables are size 5 and there are 15 cluster agents, 3 tables will be dedicated exclusively to clustering. This means that no cluster agents will ever meet non-cluster agents, severely limiting the number of meetings achievable; even extending to 4 tables would be a significant relaxation. This also limits representation; for example, if all 15 cluster agents are also male, then balance on the cluster tables will be artificially low; the new \groupselect{} interface would normally suggest a larger number of cluster tables in this scenario. While this approach allows for valid comparisons to \legacy{}, experiments represent a lower bound on the possible performance in terms of both meetings and demographics in experiments involving clustering (3/7ths of all experiments). We also chose $n_S=5$ rounds of swaps: in the most complex case, this resolved in a runtime of $<$10 seconds, so better performance is likely possible with further rounds of swaps. Finally, \groupopt{} implementations used a flat $p_s=0.5$ parameter for mixing between meetings and demographic prioritisation. A different value could improve meetings at the expense of diversity, or vice versa, even more than the below: so again, results are a lower bound on the possible performance for either meetings or demographics.

We note that the our experimental variables range from binary (two possible values) to five possible values. It has been observed in previous implementations of \legacy{} that the algorithm has struggled to allocate data sets with many-option variables, for example participants' granular states or cantons. We truncate our experiments at five options for two reasons. The maximum is as high as five to show that the new approach can handle variables with more than two or three options. The maximum is as low as five because more options than this does not make sense as input for the algorithm: if a variable has levels that are granular and only held by a small subset of participants, it is objectively difficult to fairly allocate these between tables (i.e. how should the algorithm allocate the 3 participants from city X across 10 tables). We recommend such variables are pre-processed into broader groups (e.g. region). The \groupselect{} interface is being updated to automatically suggest where and how variables levels should be clustered before running the algorithm.  

\subsection{Evaluating the Algorithm}
We require metrics to compare results across the two algorithms:
\begin{itemize}
    \item Meetings: influenced by the results in Barrett et al \cite{Barrett2023NowOptimization}, we use a geometric saturation function $f$ to evaluate algorithmic performance. The function $f$ evaluates the value to the process of the number of times each pair meets. We wish this function to be increasing (i.e. meetings add value) and concave (i.e. a pair's second meeting is less valuable than their first, as they have already exchanged ideas). Among the candidates for a `good' saturation function, there are two promising families. The first is harmonic, where each pair's $n^{th}$ meeting generates a marginal utility of $1/n$. This leads to a slow decrease in marginal utility, placing significant value on later meetings. However, it leads to strange comparative behaviour: two pairs meeting for a second time ($2/2$) generate the same utility as a single pair meeting for the first time (1), but two pairs meeting for the third time ($2/3$) generate significantly more utility than a pair meeting for the second time ($1/2$). We therefore turn to the geometric saturation function, where each pair's $n^{th}$ meeting generates marginal utility of $a^n$ for some constant $0<a<1$ (we use $a=0.5$). This leads to a faster decrease in marginal utility than the harmonic function, but has the appealing `self-similarity' property whereby once everyone has met $x$ times, the problem effectively resets (as the $x+1^{st}$ meeting is worth precisely $a$ $x^{th}$ meetings, which is the same as a second meeting being worth $a$ first meetings). In our evaluation, each pair's total meeting score over the process is aggregated to get a direct comparison of meeting qualities between \legacy{} and \groupopt{}
    \item Demographics: For each diversification demographic, in each round and on each table, we calculate the distance from the ideal balance on the table, and these are averaged into the value $\bar\Delta^A$ for allocation $A$. If these distance scores are different between algorithms for a demographic, then the algorithm with the lower score is on average closer to satisfying balance criteria for the demographic
\end{itemize}

We also derive a secondary meetings metric $excess$, used for comparing outcomes with the optimal possible performance. The optimal round-by-round performance is derived in Barrett et al \cite{Barrett2023NowOptimization}. Consequently, we can derive the number of pairs who haven't met after round $\roundno$, which we denote $N^\roundno(0)$. There are $N^P=0.5*\mid\nagents\mid * (\mid\nagents\mid - 1)$ total pairs in the panel. In each round, $M^\roundno$ pairs meet, which is a function of the space constraints, with tables sizes $\{N_{lower},N_{upper}\}$:

\begin{align*} 
M^\roundno=0.5* (Z_{l}*N_{lower}*(N_{lower}-1) + Z_{u}*N_{upper}*(N_{upper}-1))\\
Z_{l} = \mid\ntables\mid - (\mid\nagents\mid \% \mid\ntables\mid)\\
Z_{u} = \mid\nagents\mid \% \mid\ntables\mid\\
\end{align*}

In each round subsequent to the first, at least $L_R^*(\roundno,\roundno')$ pairs have a repeated meeting, so the number of pairs who haven't met is bounded by: $$N^\nrounds(0)\geq N^P - (M^\roundno + (\nrounds - 1) * (M^\roundno - L_R^*(\roundno,\roundno'))$$
In the optimal setting where this equality is achieved, the number of pairs who haven't met is denoted $N^*(0)$; actual algorithm performance will lead to an excess number of pairs yet to meet, which we report as a percentage of all pairs:

$$excess = \frac{N(0)-N^*(0)}{N^P}$$

We note that algorithmic performance is less likely to approach this bound in two cases: where additional constraints make the round-wise bound difficult to attain, and thus very difficult to attain multiple times, and in the case where all agents meet and there are no more first meetings to generate. To make the bounds calculated in $L^*_R$ meaningful, we restrict analysis of $excess$ to experiments without clustering constraints.

\section{Results}
\label{app:results}
We present the set-up and results from 66 experiments, each run for both \legacy{} and \groupopt{}.

\subsection{Experiment Outcomes}
For every single experiment, the meeting score is higher under \groupopt{} than \legacy{}. This difference becomes more significant for longer runs, as shown in Table \ref{tab:implementation_meeting_score}.

\begin{center}
\captionof{table}{Average difference in the meeting score between \legacy{} and \groupopt{} across experiments for data with $ \mid\nagents\mid$ agents and $\nrounds$ rounds}
\begin{tabular}{ |c|c|c|c|c|c| } 
 \hline
 $\nrounds$ & $ \mid\nagents\mid=30$ & $ \mid\nagents\mid=50$ & $ \mid\nagents\mid=100$ & $\mid\nagents\mid=120$\\
 \hline
 3 & 9.6 & 44.1 & 82.7 & 3.0\\ 
 5 & 28.6 & 117.0 & 261.5 & 51.9\\ 
 10 & 68.2 & 272.5 & 855.7 & 149.9\\ 
 \hline
\end{tabular}
\label{tab:implementation_meeting_score}
\end{center}

\begin{figure}[H]
\centering
\begin{subfigure}{.5\textwidth}
  \centering
  \includegraphics[width=\linewidth]{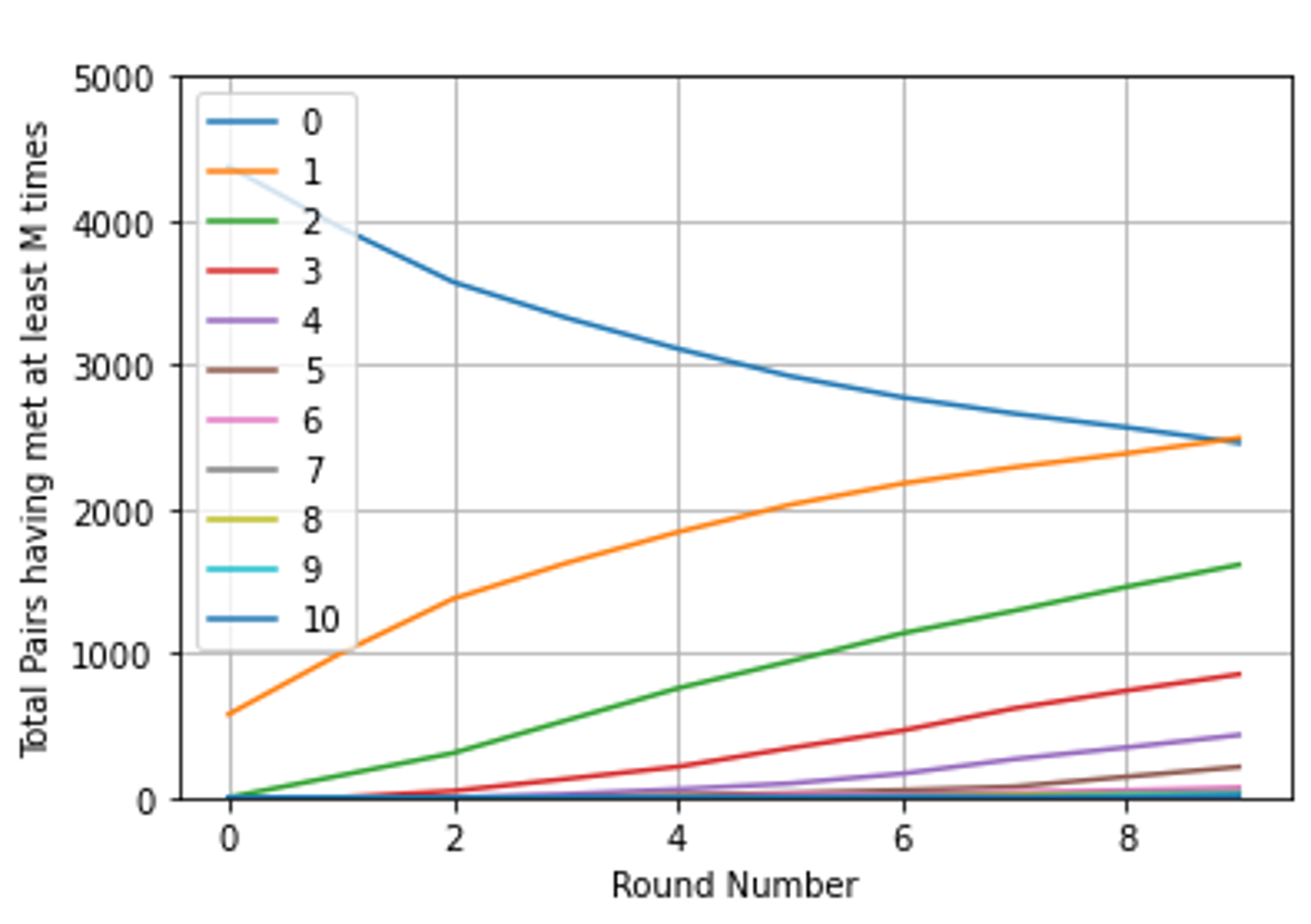}
  \caption{\legacy{}}
  \label{fig:meetings_GS}
\end{subfigure}%
\begin{subfigure}{.5\textwidth}
  \centering
  \includegraphics[width=\linewidth]{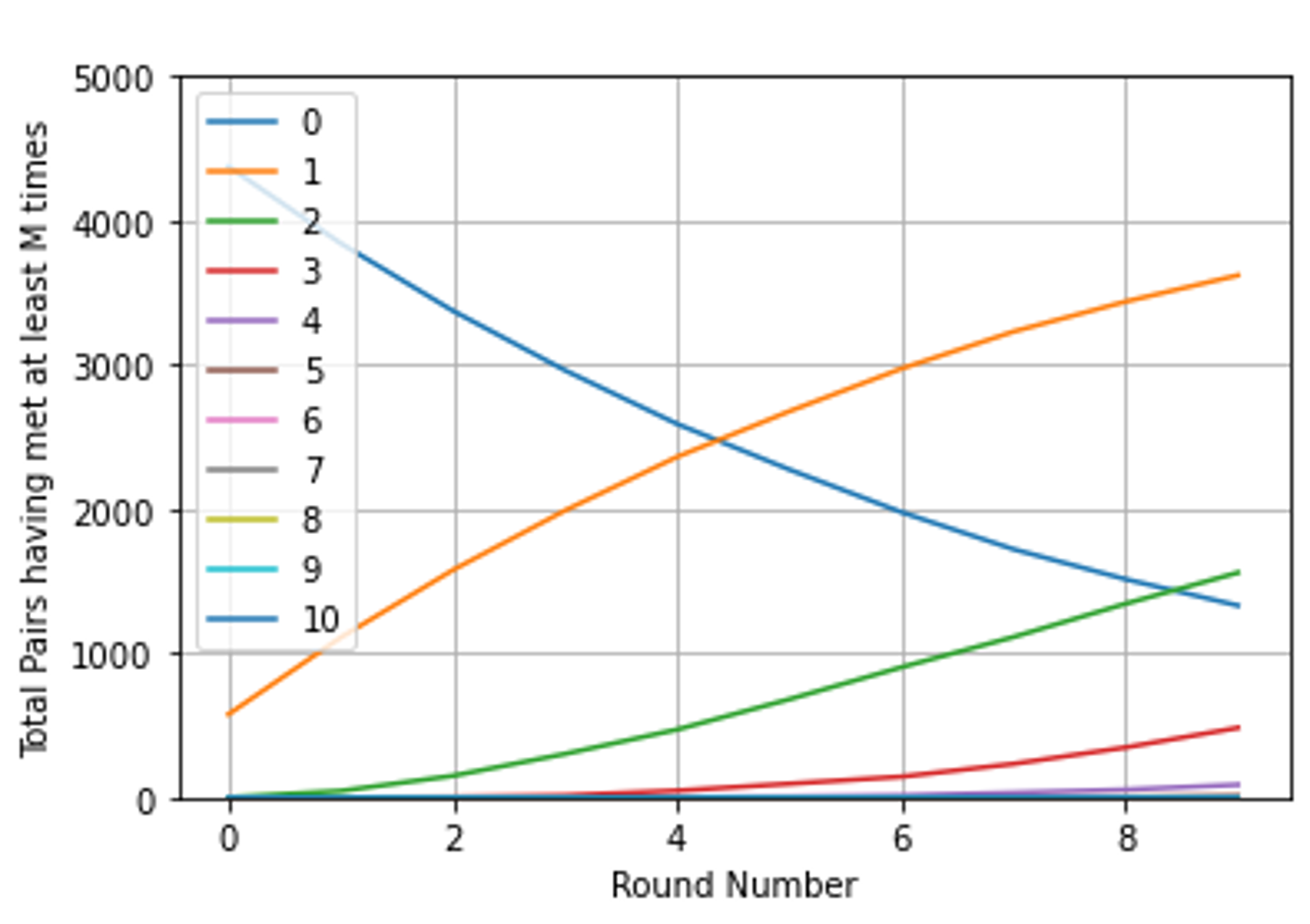}
  \caption{\groupopt{}}
  \label{fig:meetings_GB}
\end{subfigure}
\caption{Unique meetings for the $ \mid\nagents\mid=100$ data set, over $\roundno=10$ rounds. The downward-sloping line labelled ``0'' is a count of how many unique pairs haven't met in round $\roundno$, while all other lines labelled ``$M$'' are cumulative counts of how many pairs have met at least $M$ times in round $\roundno$}
\label{fig:test}
\end{figure}

Thus, performance improves at a higher rate when the algorithm runs for longer time periods. We also note that, aside from the artificially constrained $\mid\nagents\mid=120$ case with large groups, more participants with a fixed table size also leads to a greater performance increase for \groupopt{} over \legacy{}. To give an illustration of how this works, we can focus on a specific experiment for the $ \mid\nagents\mid=100$, $\nrounds=10$ rounds example in Figures \ref{fig:meetings_GS} and \ref{fig:meetings_GB}.

 Following the approach in Barrett et al \cite{Barrett2023NowOptimization}, we can also see how our \groupopt{} algorithm compares to \legacy{} when we judge it by its own performance criteria: maximising the number of first meetings. Based on previous calculations of the minimum possible number of repeated meetings between any two rounds, we can calculate the maximum number of first meetings that are possible, given the panel size and number of tables.\footnote{Note that in many cases this number will not be achievable, as reallocating in a way that generates the minimum number of repeated meetings may only be possible once} Under this approach, \legacy{} achieves 62.6\% of possible first meetings averaged over non-clustering implementations. \groupopt{} achieves 76.9\% of all possible first meetings.

When considering optimal performance metric $excess$, \groupopt{} outperforms \legacy{} on all but one of the experiments. The exception is a simple case with $\nagents=30$, $\ntables = 2$, $\nrounds=3$, where \groupopt{} leads to three fewer pairs meeting than \legacy{} (a proportionate difference of 0.7\% of all pairs). Across all experiments, the average improvement generated by \groupopt{} (i.e. $excess^{\legacy{}}-excess^{\groupopt{}}$) is 12.7\%.

Differences in average diversity scores $\bar\Delta^{A,\legacy{}}-\bar\Delta^{A,\groupopt{}}$ are shown in Figure \ref{fig:diversity}:

\begin{figure}[H]
\centering
\includegraphics[width=0.6\textwidth]{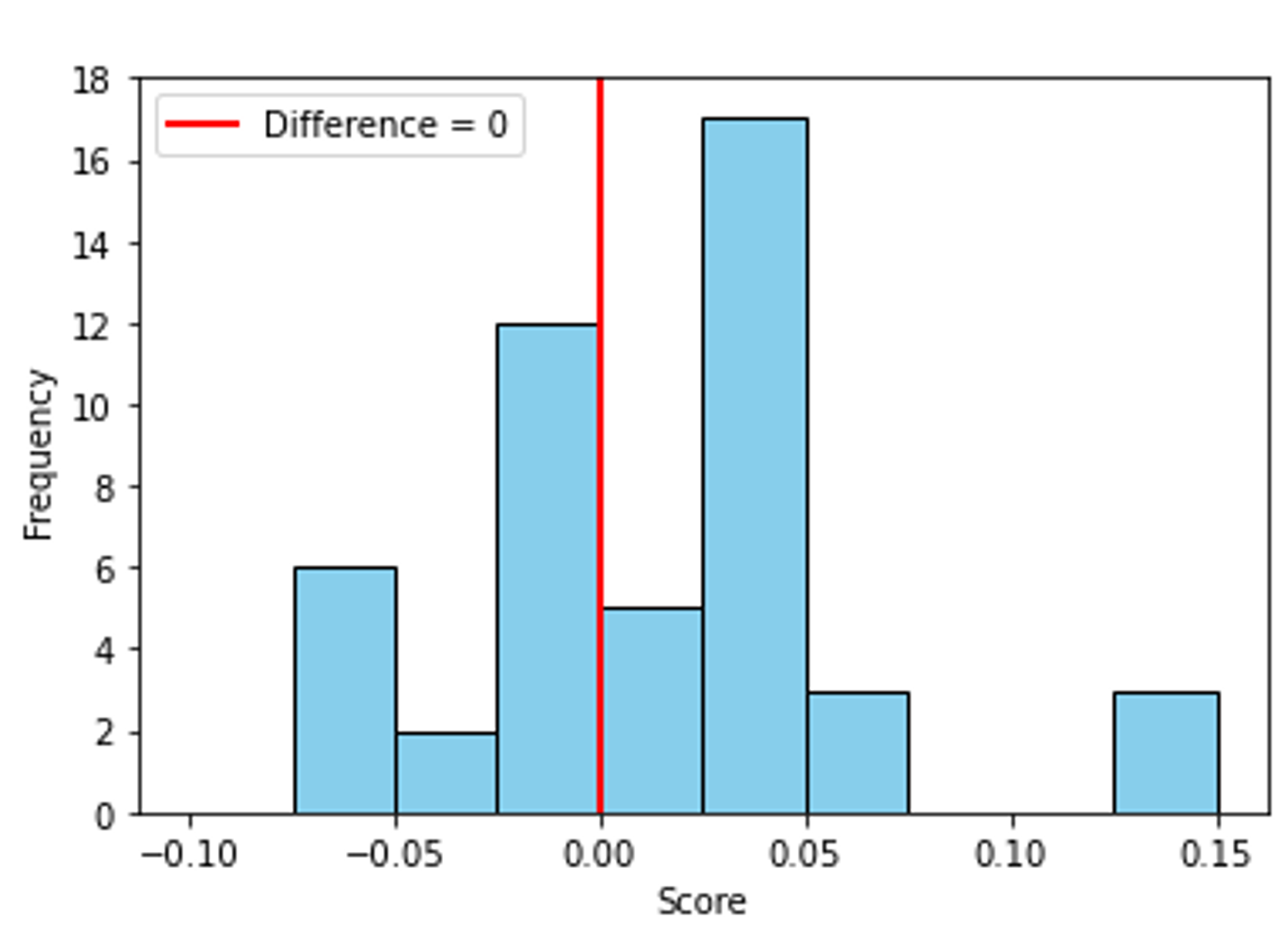}
\caption{Histogram of experiments in which \groupopt{} attains better (positive values), the same (0 values) or worse (negative values) diversity scores than \legacy{}}
\label{fig:diversity}
\end{figure}

Note the cluster of experiments with value 0, denoted by a red line: these are straightforward implementations that are analytically easy to solve (i.e. few demographics and low $ \mid\nagents\mid$). Although results skew in favour of \groupopt{}, 20 experiments have worse diversity scores than \legacy{} implementations. This is largely due to the artificially tight constraints imposed on \groupopt{} to match the experiment set-up of \legacy{}, meaning that these results are a significant lower bound. These clustering constraints are particularly tight in smaller settings e.g. low numbers of tables and participants. Larger $ \mid\nagents\mid$ and no clustering are associated with the better balance outcomes. A higher value of $p_S$ ensures that Pareto swaps are more likely to prioritise demographic scores over meeting scores; as meeting scores are so much better than \legacy{}, there is wide scope to do this for end users. Note that \legacy{} places an importance ordering on variables, whereas \groupopt{} tries to balance all variables. As such, \legacy{} is sometimes slightly better at balancing the earliest 1 or 2 variables, but much worse at balancing later variables.

Finally, we note that \groupopt{} is trivially better than random allocation by construction, as the initial allocation is based on randomness and the set of Pareto swaps explicitly improve upon this.

\subsection{Comparisons with Optimal}
\label{app:optimal}
The attentive reader may note that we have omitted direct comparisons with the algorithm from Barrett et al \cite{Barrett2023NowOptimization}. This is because their approach only seeks optimal allocations on a greedy round-by-round basis, and therefore does not represent the best possible performance. Having established that \groupopt{} outperforms \legacy{}, in this Section we explore how close the algorithm gets to true optimum. 

For demographics, the true optimum is exact demographic balance for each demographic on each table in each round, to the nearest whole number. In all experiments, every single table is within 10\% of the optimal percentage balance for every demographic considered. For larger datasets, where this effectively represents a single participant, this represents as close to optimal as possible given the space constraints. 

\begin{figure}[H]
\centering
\includegraphics[width=0.6\linewidth]{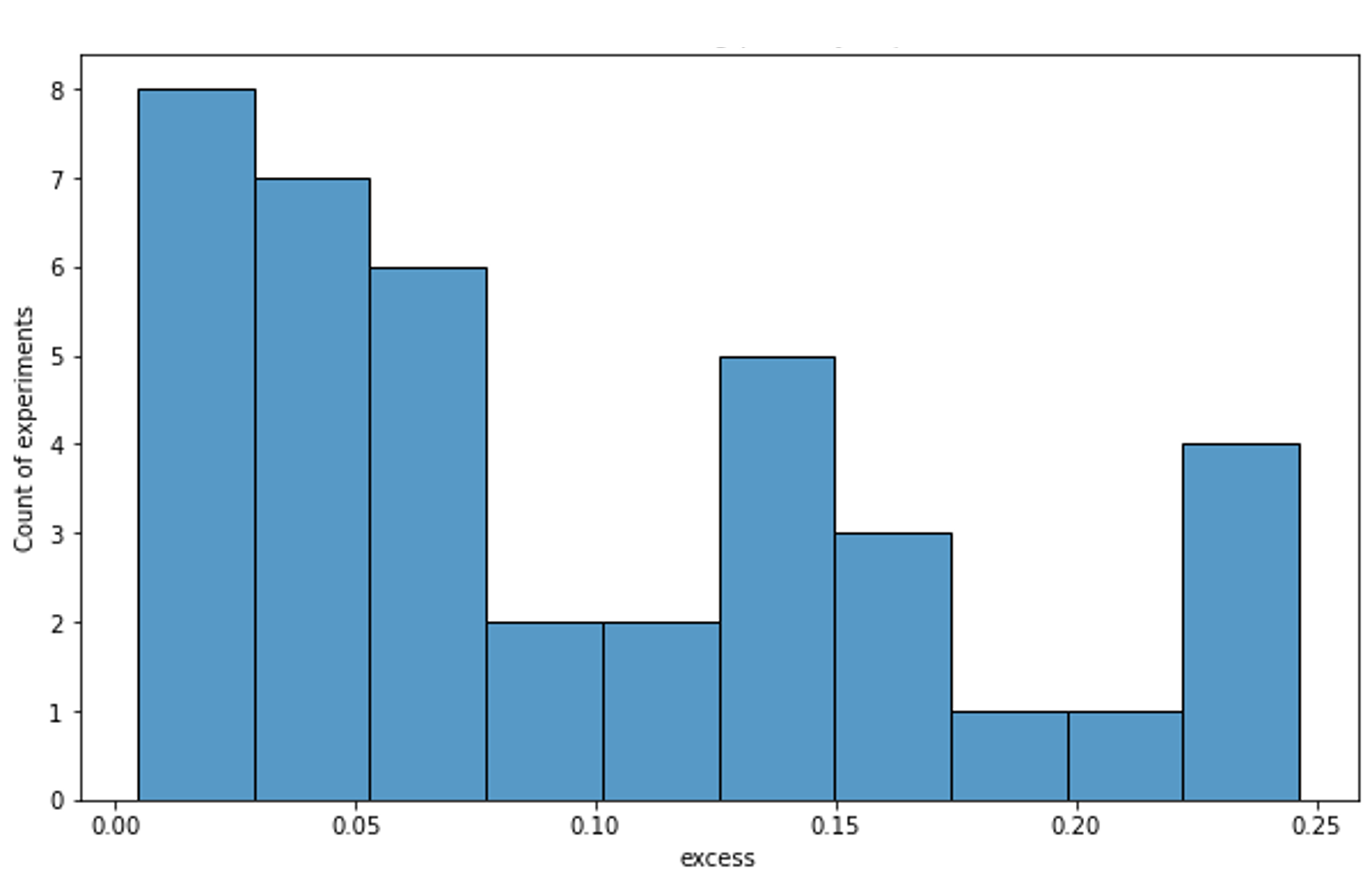}
\caption{$excess$ values for \groupopt{} experiments: the proportion of pairs that don't meet over the panel but theoretically could}
\label{fig:excess}
\end{figure}

The excess number of pairs who do not meet under \groupopt{} (Figure \ref{fig:excess}) ranges from a minimum of 0.5\% of all pairs to a maximum of 24.6\%, with an average score of 9.5\%. The cluster of results with an excess of under 10\% (59.0\% of experiments) indicates that the algorithm performs close to optimal in most settings, while the relatively low maximum value indicates that even in complex situations with multiple diversification variables, performance does not leak to a dramatic extent. For context, the worst-performing experiment for \legacy{} sees 74.0\% of pairs not meeting who theoretically could.

\section{Future Developments}

Given the highly promising experimental results, we are currently working on implementing the \groupopt{} algorithm and an updated \groupselect{} interface with the Sortition Foundation. Beyond implementation, these results leave open questions. The persistence of some excess meetings over an optimal performance is partially a function of the problem setting: the possibility of achieving the minimum $L_R^*(\roundno,\roundno')$ repeated meetings between two rounds does not necessarily imply that this result scales for 3+ rounds, as the problem becomes exponentially more complex with more rounds. However, the higher $excess$ in the worse-performing experiments indicates that some further improvement on \groupopt{} is likely attainable. Additionally, a user study on how practitioners deploy this algorithm in the field is vital to ensuring that both the algorithm is used as intended and the front-end features, such as app design and layout, are optimised for users' convenience.

\bibliography{references}


\begin{thebibliography}{12}
\ifx \bisbn   \undefined \def \bisbn  #1{ISBN #1}\fi
\ifx \binits  \undefined \def \binits#1{#1}\fi
\ifx \bauthor  \undefined \def \bauthor#1{#1}\fi
\ifx \batitle  \undefined \def \batitle#1{#1}\fi
\ifx \bjtitle  \undefined \def \bjtitle#1{#1}\fi
\ifx \bvolume  \undefined \def \bvolume#1{\textbf{#1}}\fi
\ifx \byear  \undefined \def \byear#1{#1}\fi
\ifx \bissue  \undefined \def \bissue#1{#1}\fi
\ifx \bfpage  \undefined \def \bfpage#1{#1}\fi
\ifx \blpage  \undefined \def \blpage #1{#1}\fi
\ifx \burl  \undefined \def \burl#1{\textsf{#1}}\fi
\ifx \doiurl  \undefined \def \doiurl#1{\url{https://doi.org/#1}}\fi
\ifx \betal  \undefined \def \betal{\textit{et al.}}\fi
\ifx \binstitute  \undefined \def \binstitute#1{#1}\fi
\ifx \binstitutionaled  \undefined \def \binstitutionaled#1{#1}\fi
\ifx \bctitle  \undefined \def \bctitle#1{#1}\fi
\ifx \beditor  \undefined \def \beditor#1{#1}\fi
\ifx \bpublisher  \undefined \def \bpublisher#1{#1}\fi
\ifx \bbtitle  \undefined \def \bbtitle#1{#1}\fi
\ifx \bedition  \undefined \def \bedition#1{#1}\fi
\ifx \bseriesno  \undefined \def \bseriesno#1{#1}\fi
\ifx \blocation  \undefined \def \blocation#1{#1}\fi
\ifx \bsertitle  \undefined \def \bsertitle#1{#1}\fi
\ifx \bsnm \undefined \def \bsnm#1{#1}\fi
\ifx \bsuffix \undefined \def \bsuffix#1{#1}\fi
\ifx \bparticle \undefined \def \bparticle#1{#1}\fi
\ifx \barticle \undefined \def \barticle#1{#1}\fi
\bibcommenthead
\ifx \bconfdate \undefined \def \bconfdate #1{#1}\fi
\ifx \botherref \undefined \def \botherref #1{#1}\fi
\ifx \url \undefined \def \url#1{\textsf{#1}}\fi
\ifx \bchapter \undefined \def \bchapter#1{#1}\fi
\ifx \bbook \undefined \def \bbook#1{#1}\fi
\ifx \bcomment \undefined \def \bcomment#1{#1}\fi
\ifx \oauthor \undefined \def \oauthor#1{#1}\fi
\ifx \citeauthoryear \undefined \def \citeauthoryear#1{#1}\fi
\ifx \endbibitem  \undefined \def \endbibitem {}\fi
\ifx \bconflocation  \undefined \def \bconflocation#1{#1}\fi
\ifx \arxivurl  \undefined \def \arxivurl#1{\textsf{#1}}\fi
\csname PreBibitemsHook\endcsname

\bibitem[\protect\citeauthoryear{Flanigan et~al.}{2021}]{Flanigan2021FairAssemblies}
\begin{barticle}
\bauthor{\bsnm{Flanigan}, \binits{B.}},
\bauthor{\bsnm{G{\"{o}}lz}, \binits{P.}},
\bauthor{\bsnm{Gupta}, \binits{A.}},
\bauthor{\bsnm{Hennig}, \binits{B.}},
\bauthor{\bsnm{Procaccia}, \binits{A.D.}}:
\batitle{{Fair algorithms for selecting citizens’ assemblies}}.
\bjtitle{Nature}
\bvolume{596}(\bissue{7873}),
\bfpage{548}--\blpage{552}
(\byear{2021})
\doiurl{10.1038/s41586-021-03788-6}
\end{barticle}
\endbibitem

\bibitem[\protect\citeauthoryear{Flanigan et~al.}{2020}]{Flanigan2020NeutralizingSortition}
\begin{botherref}
\oauthor{\bsnm{Flanigan}, \binits{B.}},
\oauthor{\bsnm{G{\"{o}}lz}, \binits{P.}},
\oauthor{\bsnm{Gupta}, \binits{A.}},
\oauthor{\bsnm{Procaccia}, \binits{A.D.}}:
{Neutralizing self-selection bias in sampling for sortition}.
Advances in Neural Information Processing Systems
\textbf{34}
(2020)
\doiurl{10.48550/arXiv.2006.10498}
\end{botherref}
\endbibitem

\bibitem[\protect\citeauthoryear{Curato et~al.}{2023}]{Curato2023GlobalReport}
\begin{botherref}
\oauthor{\bsnm{Curato}, \binits{N.}},
\oauthor{\bsnm{Chalaye}, \binits{P.}},
\oauthor{\bsnm{Conway-Lamb}, \binits{W.}},
\oauthor{\bsnm{De~Pryck}, \binits{K.}},
\oauthor{\bsnm{Elstub}, \binits{S.}},
\oauthor{\bsnm{Mor{\'{a}}n}, \binits{A.}},
\oauthor{\bsnm{Oppold}, \binits{D.}},
\oauthor{\bsnm{Romero}, \binits{J.}},
\oauthor{\bsnm{Ross}, \binits{M.}},
\oauthor{\bsnm{Sanchez}, \binits{E.}},
\oauthor{\bsnm{Sari}, \binits{N.}},
\oauthor{\bsnm{Stasiak}, \binits{D.}},
\oauthor{\bsnm{Tilikete}, \binits{S.}},
\oauthor{\bsnm{Veloso}, \binits{L.}},
\oauthor{\bsnm{Schneidermesser}, \binits{D.}},
\oauthor{\bsnm{Werner}, \binits{H.}}:
{Global Assembly on the Climate and Ecological Crisis: Evaluation Report}.
Technical report,
University of Canberra
(2023).
\url{https://globalassembly.org/report}
\end{botherref}
\endbibitem

\bibitem[\protect\citeauthoryear{Boswell et~al.}{2023}]{Boswell2023IntegratingAssemblies}
\begin{barticle}
\bauthor{\bsnm{Boswell}, \binits{J.}},
\bauthor{\bsnm{Dean}, \binits{R.}},
\bauthor{\bsnm{Smith}, \binits{G.}}:
\batitle{{Integrating citizen deliberation into climate governance: Lessons on robust design from six climate assemblies}}.
\bjtitle{Public Administration}
\bvolume{101}(\bissue{1}),
\bfpage{182}--\blpage{200}
(\byear{2023})
\doiurl{10.1111/padm.12883}
\end{barticle}
\endbibitem

\bibitem[\protect\citeauthoryear{Andrews et~al.}{2022}]{Andrews2022ScotlandsExperience}
\begin{botherref}
\oauthor{\bsnm{Andrews}, \binits{N.}},
\oauthor{\bsnm{Elstub}, \binits{S.}},
\oauthor{\bsnm{McVean}, \binits{S.}},
\oauthor{\bsnm{Sandie}, \binits{G.}}:
{Scotland's Climate Assembly Research Report: process, impact and Assembly member experience}.
Technical report,
Scottish Government Research,
Edinburgh
(2022)
\end{botherref}
\endbibitem

\bibitem[\protect\citeauthoryear{Escobar and Elstub}{2017}]{Escobar2017FormsPractice}
\begin{botherref}
\oauthor{\bsnm{Escobar}, \binits{O.}},
\oauthor{\bsnm{Elstub}, \binits{S.}}:
{Forms of Mini-publics: An introduction to deliberative innovations in democratic practice}.
newDemocracy,
1--9
(2017)
\end{botherref}
\endbibitem

\bibitem[\protect\citeauthoryear{{OECD}}{2020}]{OECD2020InnovativeInstitutions}
\begin{botherref}
\oauthor{\bsnm{{OECD}}}:
{Innovative Citizen Participation and New Democratic Institutions}.
Technical report
(June 2020).
\doiurl{10.1787/339306} .
\url{https://www.oecd-ilibrary.org/governance/innovative-citizen-participation-and-new-democratic-institutions_339306da-en}
\end{botherref}
\endbibitem

\bibitem[\protect\citeauthoryear{{newDemocracy Foundation} and {United Nations Democracy Fund}}{2018}]{newDemocracyFoundation2018EnablingElections}
\begin{botherref}
\oauthor{\bsnm{{newDemocracy Foundation}}},
\oauthor{\bsnm{{United Nations Democracy Fund}}}:
{Enabling National Initiatives to Take Democracy Beyond Elections}.
Technical report
(2018)
\end{botherref}
\endbibitem

\bibitem[\protect\citeauthoryear{Elstub et~al.}{2022}]{Elstub2022ResearchScotland}
\begin{botherref}
\oauthor{\bsnm{Elstub}, \binits{S.}},
\oauthor{\bsnm{Escobar}, \binits{O.}},
\oauthor{\bsnm{Henderson}, \binits{A.}},
\oauthor{\bsnm{Thorne}, \binits{T.}},
\oauthor{\bsnm{Bland}, \binits{N.}},
\oauthor{\bsnm{Bowes}, \binits{E.}}:
{Research Report on the Citizens' Assembly of Scotland}.
Technical report,
Scottish Government
(2022).
\url{https://www.gov.scot/isbn/9781802018943}
\end{botherref}
\endbibitem

\bibitem[\protect\citeauthoryear{Barrett et~al.}{2023}]{Barrett2023NowOptimization}
\begin{bchapter}
\bauthor{\bsnm{Barrett}, \binits{J.}},
\bauthor{\bsnm{Gal}, \binits{Y.K.}},
\bauthor{\bsnm{G{\"{o}}lz}, \binits{P.}},
\bauthor{\bsnm{Hong}, \binits{R.}},
\bauthor{\bsnm{Procaccia}, \binits{A.D.}}:
\bctitle{{Now we're talking: Better deliberation groups through submodular optimization}}.
In: \bbtitle{Proceedings of the 37th AAAI Conference on Artificial Intelligence},
pp. \bfpage{5490}--\blpage{5498}
(\byear{2023}).
\doiurl{10.1609/aaai.v37i5.25682}
\end{bchapter}
\endbibitem

\bibitem[\protect\citeauthoryear{Huangfu and Hall}{2018}]{Huangfu2018ParallelizingMethod}
\begin{barticle}
\bauthor{\bsnm{Huangfu}, \binits{Q.}},
\bauthor{\bsnm{Hall}, \binits{J.A.J.}}:
\batitle{{Parallelizing the dual revised simplex method}}.
\bjtitle{Mathematical Programming Computation}
\bvolume{10}(\bissue{1}),
\bfpage{119}--\blpage{142}
(\byear{2018})
\doiurl{10.1007/s12532-017-0130-5}
\end{barticle}
\endbibitem

\bibitem[\protect\citeauthoryear{{Sortition Foundation}}{2021}]{SortitionFoundation2021GroupAlgorithm}
\begin{botherref}
\oauthor{\bsnm{{Sortition Foundation}}}:
{Group Select Algorithm}
(2021).
\url{https://www.sortitionfoundation.org/services}
\end{botherref}
\endbibitem

\end{thebibliography}


\end{document}